\documentclass[sn-mathphys]{sn-jnl}% Math and Physical Sciences Reference Style
\usepackage{natbib}
\setcitestyle{open={(},close={)}} %Citation-related commands

\jyear{2021}%

\theoremstyle{thmstyleone}%
\theoremstyle{thmstyletwo}%

\theoremstyle{thmstylethree}%

\begin{document}

\title[Use of Topical and Temporal Profiles and their Hybridisation for CBR]{Use of Topical and Temporal Profiles and their Hybridisation for Content-based Recommendation\footnote{This paper or a similar version is not currently under review by a journal or conference. This paper is void of plagiarism or self-plagiarism as defined by the Committee on Publication Ethics and Springer Guidelines}}

%%=============================================================%%
%% Prefix	-> \pfx{Dr}
%% GivenName	-> \fnm{Joergen W.}
%% Particle	-> \spfx{van der} -> surname prefix
%% FamilyName	-> \sur{Ploeg}
%% Suffix	-> \sfx{IV}
%% NatureName	-> \tanm{Poet Laureate} -> Title after name
%% Degrees	-> \dgr{MSc, PhD}
%% \author*[1,2]{\pfx{Dr} \fnm{Joergen W.} \spfx{van der} \sur{Ploeg} \sfx{IV} \tanm{Poet Laureate} 
%%                 \dgr{MSc, PhD}}\email{iauthor@gmail.com}
%%=============================================================%%

\author[1]{\fnm{Luis M.} \sur{de Campos}}\email{lci@decsai.ugr.es}

\author*[1]{\fnm{Juan M.} \sur{Fernández-Luna}}\email{jmfluna@decsai.ugr.es}

\author[1]{\fnm{Juan F.} \sur{Huete}}\email{jhg@decsai.ugr.es}
\equalcont{These authors contributed equally to this work.}

\affil*[1]{\orgdiv{Departmento de Ciencias de la Computación e Inteligencia Artificial}, \orgname{Escuela Técnica Superior de Ingenierías Infomática y de Telecomunicación, CITIC-UGR, Universidad de Granada}, \orgaddress{\street{C/ Periodista Daniel Saucedo Aranda, S/N}, \city{Granada}, \postcode{18013}, \country{Spain}}}

%%==================================%%
%% sample for unstructured abstract %%
%%==================================%%

\abstract{In the context of content-based recommender systems, the aim of this paper is to determine how better profiles can be built and how these affect the recommendation process based on the incorporation of temporality, i.e. the inclusion of time in the recommendation process, and topicality, i.e. the representation of texts associated with users and items using topics and their combination. The main contribution of the paper is to present two different ways of hybridising these two dimensions and to evaluate and compare them with other alternatives.}

\keywords{Content-based recommendation, topical profiles, temporal profiles, publication venue recommendation}

%%\pacs[JEL Classification]{D8, H51}

%%\pacs[MSC Classification]{35A01, 65L10, 65L12, 65L20, 65L70}

\maketitle

\section{Introduction}

In today's world, recommender systems (RSs) \cite{Bobadilla13} play an extremely important role in our digital life \cite{LWMWZ15}. Many e-commerce platforms (such as Amazon or eBay, for example), entertainment services (e.g. Netflix, Spotify), or social media (Instagram, Facebook, Twitter, etc.) incorporate recommendation into their functionalities. These systems analyse how users interact with their products and suggest ones that the users might be interested in.

There are two main types of recommender systems \cite{AT05}: those based on collaborative filtering (CF) and those based on content-based recommendation (CBR) \cite{Lops19}. While the first type generates suggestions based on user ratings for items they have consumed, the second offers recommendations based on the textual contents of these consumed items. In both cases, ratings or contents are stored in special structures called profiles \cite{Gauch07} and these represent the users' interests. RSs exploit these profiles to generate useful suggestions for users.

If we focus on CBR, as long as there is a textual description of the items, users' profiles are usually represented by a bag of words which is obtained by combining the texts for any item the user has shown interest in. As CBR is primarily based on information retrieval (IR), the recommendation process consists in carrying out a matching between the active user's profile (in terms of IR, a representation of an information need, i.e. a query) and the textual representation of all the items in the collection (documents in the IR context). The degree of relevance of each item with respect to the corresponding profile is therefore computed and a ranking of relevant items to be recommended is subsequently generated.

This approach is used for most recommendation contexts (books, films, music, etc.). There are, however, other situations where items are associated with a group of documents rather than a single text, such as expert finding \cite{Lin17}, whereby experts on a certain subject are recommended according to a set of documents that define their expertise to any user who requires them, or publication venue recommendation \cite{Wang18}, where journals or conferences are recommended to scientists who wish to know where to publish a paper. In this case, the venues comprise a series of articles published there. In these two cases, item profiles might be built to define their specific informational context and subsequently used in the recommendation stage. This context in which the items are also described by means of profiles compiled from the associated documents represents the general premise of this paper.

As we have already mentioned, a large proportion of the user profile consists of (weighted) terms or keywords, i.e. the words included in the textual descriptions of items associated to a number that reflects the importance of each term in the item and/or in the entire collection of terms. Other alternatives have, however, been explored in the literature in terms of profile construction such as the inclusion of topical and temporal dimensions. 

Regarding such topic-based profiles, in a first approximation, these comprise topics, concepts or categories rather than terms. They may be easily constructed by including the topics of interests or categories of an item (for example, in a recommender system for the digital edition of a journal or magazine, categories such as sport, local or finances, for example, are associated with each news item, and these may be directly moved to its profile). In a second approach, however, the items could be represented by several subprofiles rather than a single one. In this case, textual subprofiles could be created for each topic, incorporating the text of all the associated documents about this topic. Let us imagine a simple situation in the context of expert finding in academia. Let us suppose that an expert, a researcher, has published articles about RS, IR, personalisation and applications. The researcher's profile would consequently consist of four subprofiles, each containing the texts of the articles in that area of expertise. Each item, therefore, would be represented by several subprofiles that are more topically homogeneous, as documents are not mixed in one profile (heterogeneous) but grouped according to topic of interest. This method of organisation would clearly increase the interpretability of these (sub)profiles. This topical dimension can be automatically incorporated to the profiles by mining the texts in situations in which these categories are not clearly available, such as, for example, learning topic models using Latent Dirichlet al.location (LDA) \cite{deCampos21,JWY19} or by applying clustering algorithms \cite{deCampos20}.

In situations where time is a feature included in those documents that define items (such as any type of timestamp), this dimension could be incorporated in different ways into the recommendation \cite{CDC14}. Firstly, temporal subprofiles may be built by dividing the temporal line of documents into periods, grouping them into each one of these and thereby building the corresponding subprofiles. The homogeneity of the profiles (in this case temporal) is also present in the item profiling process. This approach could be considered as a generalisation of the well-known profiles based on long and short-term preferences. Secondly, but unrelated to profiles, another option is to include time in the recommendation process, by means of the application of a decay factor that penalises older items.

Bearing in mind these topical and temporal dimensions for building homogeneous subprofiles, a question immediately arises: would it make sense to combine both aspects in order to improve the quality of the recommendation in terms of system effectiveness? Since the state of the art shows that it is suitable, two innovative methods for carrying out such an integration are presented and evaluated in this paper and these represent the main contribution of this article. Our goal is, therefore, to determine whether this mixture is valuable in comparison with other non-hybrid alternatives. For this purpose, this study will address the following research questions:

\begin{itemize}
    \item RQ1: Does a temporal division of the documents associated to items provide a reliable source for constructing high quality profiles? 
    \item RQ2: Can decay-based techniques which penalise older documents be successfully incorporated?
    \item RQ3: To what extent is building item profiles based on latent topics in the document collection an added valued for the recommendation problem?
    \item RQ4: Is hybridisation, i.e. the combination of topical and temporal aspects for creating item profiles, a good alternative for the problem at hand?
    \item RQ5: Which method is the best form of hybridisation?
\end{itemize}

Although the profiling proposals presented in this paper may be applied to any type of item collection represented by text in the context of CBR, evaluation is performed in the field of publication venue recommendation, i.e. given a paper which is about to be submitted to a certain venue for revision, the recommender system would suggest the most suitable journals, conferences or general scientific events based on the subject of the paper. The CBR system will build subprofiles for venues, starting from the collection of published articles, which will be matched against the text of the submitted article. 

The remainder of this paper is organised as follows: Section \ref{relatedwork} briefly discusses other related works; Section \ref{alternatives} introduces the different methods for creating (sub)profiles based on terms, topics and time, and for combining them; Section \ref{evaluation} focuses on the experimental part of the work, including results and discussion; and, finally, Section \ref{conclusions} details our conclusions and outlines our future lines of research.

\section{Related Work} \label{relatedwork}

Starting with the simplest form of representing items or user profiles, the method usually adopted is to compile a list of weighted terms which are automatically extracted from the document associated with them \cite{Gauch07}. These terms are supposed to correctly represent the document subjects and the weights are responsible for measuring their importance in terms of the entire document collection and within each document. Some examples of the use of term-based profiles for recommendation reflect how widely they are used in CBR and these include TV programme recommendation \cite{WS11}, expert finding \cite{deCampos20}, treatments for patients in a health RS \cite{BDB18}, tweet recommendation \cite{BF16}, and image recommendation \cite{KEA18}.

An alternative way of representing profiles to keywords is through the use of tags, concepts, categories or topics. We could say that these are higher-level features which could symbolise concepts and try to capture the underlying semantics of the items. Since the profiles comprise more general concepts rather than just words, certain authors believe that this is beneficial for the quality of the recommendation \cite{FWP07}, at least in the case of tags. There are a number of examples of experiences that build and recommend based on tag profiles \cite{Boger18,AC18,SH19,YCHZL20,BGG17}. Tags are assigned to user profiles either manually or on the basis of a machine learning-based approach. The use of concepts is discussed in a number of papers \cite{RZCDS15,NBM16,SGM17,SK18}. In most cases, the concepts are extracted from ontologies or concept graphs giving the information associated to items or users. With respect to topic-based profiles \cite{SCS20}, these comprise latent topics mined directly from document collections, typically using the LDA algorithm or extensions of it. Starting with user-associated texts, the profile is fed with the most probable topics associated to the words contained in them. Certain papers illustrate the use of topic-based profiles to CBR in a wide variety of problems \cite{CMXL17,HW19,deCampos21,FSCH21,KSC21}. Although the underlying representation based on topics is very appropriate for representing profiles, it does, in fact, lack the interpretability offered by terms, tags or concepts. A translation from topics to human-understandable labels is needed and this requires an additional effort. 

All these previous item or user profiles are monolithic in the sense that all the possible facets of interests are combined into a single profile. Another alternative is to consider profiles as comprising different subprofiles, each associated to a possible facet, thereby capturing the various underlying, non-explicit topics which are usually extracted by machine learning algorithms from the associated texts of users and items. These multi-faceted profiles are no longer flat although they may have different shapes: trees, representing personal data, expertise and interests \cite{PL15}; graphs of clusters capturing different facets from different sources \cite{ZCM02}; two subprofiles to capture user interests and friends' interests \cite{GFSC14}; subprofiles comprising subsets of items rated by the user and which are used to improve the diversity of the recommendations \cite{Kaya19}; different types of subprofiles, each containing keywords, concepts and tags \cite{NMSLG13}, or hierarchies of weighted topics \cite{Koo05}. Clustering is the usual technique for creating such multi-faceted profiles. This unsupervised learning is applied to the document collection resulting in clusters of documents or keywords, which will integrate the profiles as subprofiles. Each cluster would represent a concept in the entire collection. There are a number of papers which cover this methodology  \cite{SH01,MKS02,YGS09,AIOS14,deCampos20,deCampos21}.

In the research presented in this paper, topic-based profiles will comprise subprofiles which represent different concepts but rather than containing a list of topics, they contain terms, i.e. those from the documents associated with the topics. 

Much has been published on temporal dynamics, i.e. the inclusion of time in recommendation, and this has mainly focused on CF \cite{CDC14}. One of the most common approaches is to use decay functions to penalise old items and reward new ones \cite{Ding05,Yeniterzi15}. A second alternative is to include time in the computation of item weights \cite{LB19}. Another possibility is to integrate time into the rating matrix in CF and use it to find trusted relationships between users \cite{NAC21}. Another research line is to consider time frames: in the article \citet{Ramos14}, the authors propose a CBR system for tweets, where a specific time frame is learned for each user and only tweets within this personalised frame are recommended. The same idea has been used by other authors \cite{SZL17} but for points-of-interest recommendation. One generalisation is the use of long and short-term profiles as another option to include time and take into account the users' most recent interests in contrast to those which were acquired by interacting with the system some time ago \cite{LZL11,XiangEtAl10}. The time domain is included in our models simply by splitting the documents into time periods of equal size rather than using long and short-term profiles. Within each time period, the topic subprofiles are learnt.

Finally, this review of related work will examine the combination of topicality and temporality in profiles by taking advantage of both dimensions in order to improve recommendation. This combination is performed by following a wide range of methods which are outlined in the papers mentioned, but the most usual way is to apply a latent topic discovery algorithm to the available text collections, obtain the topics associated to each document and incorporate time by means of weights associated to topics. Other authors use decay functions \cite{WW21} to mitigate the impact of old ratings. They also use item reviews in order to obtain the underlying topics in the collection and associate the rated items to the corresponding topics in the reviews in order to track how the topics evolve with time. Finally, they propose an optimisation method to make predictions. In the article \cite{LZYL14} about news recommendation, the authors build long and short-term profiles. While long-term profiles comprise latent topics extracted from LDA from a collection of news and weighted by considering a time decay function to capture how they evolve in time, short-term profiles comprise topics occurring in documents in the most recent period of time. In the article \cite{YCCHZ15}, their authors  describe a method for the context of social media to combine interests and temporal context. It is based on mixing a latent class statistical mixture model to represent topic distributions not only from users' interests (user-oriented topics) but also from a temporal context (time-oriented topics). They also compute the distribution of topics for items. With all of this information, they are able to model different users' interests in different time periods. In \cite{Liu15}, it is considered the interaction of each user $u$ with each item $i$ in a given time. LDA is applied to extract topics from the set of textual representations of all these interactions, which are represented by a topic distribution. For a given user, once all of their interactions have been sorted chronologically, the assigned topics are modelled as a time series. Recommendation takes place when a Gaussian process predicts the value of each topic at a given time and similarities are computed between the predicted topic distribution and the distribution associated with each user. In another article \cite{NFB17} about community question answering, the authors introduce a method for future expert finding which suggests the most suitable experts for the future. In order to do so, they first apply LDA in order to extract topics from documents associated to experts, and their corresponding timestamps, and calculate the probability of a future expert candidate for a given query. In the context of social media \cite{NS16}, users' interests are extracted from social media streams. Profiles are then built using weighted concepts, which are the topics obtained by applying LDA to the collection. Items are also indexed using concepts and matched to user profiles. Recommendation is carried out by computing a similarity between user profiles and item profiles. Publication times are also taken into account by means of decay functions, which penalise the older topics and are included in the topic weights. In \cite{ZDZYD18}, it is built temporal user profiles by directly incorporating time into the LDA algorithm, thereby obtaining topic distributions for words and times, as in the case of timeSVD++ is \cite{Koren10}. The last two papers on expert finding use concepts rather than topics from LDA. In \cite{RBN14}, the profiles consist of weighted concepts, where the weights represent the degree of expertise in each concept. The temporal expertise profile is a set of single profiles which are computed at different periods of time, and a decay function is incorporated into the calculation of the concept weights. In \cite{ZGBBH12}, while short-term profiles are built by extracting and weighting concepts from an ontology over given time periods, long-term profiles are built by detecting the concepts which are uniformly distributed in the short-term profiles. 

The way in which temporality and topicality are combined in this paper is a contribution to the state of the art. In most cases where LDA is used, it has been applied globally to the entire document collection. In this research, LDA is applied locally only to those documents belonging to the same period of time when both dimensions are combined.

\section{Alternatives to profile construction based on terms, topics and time} \label{alternatives}
\subsection{Term-based Profiles}

In this paper, and in the context of CBR, we assume that each item to be recommended has an associated set of text documents. For example, in terms of the context of our experimentation in this paper, we are dealing with journal recommendation, where the journals are the items and all the articles published in each one are linked to it. The active user, in this case a researcher, would be interested in knowing possible journals where they could publish a recently written paper. A second example might be the so-called Expert Finding area \cite{Balog12,Lin17}. In this case, the ``items'' are experts in an area (for instance, researchers, lawyers or politicians) and their documents are all the possible associated texts (e.g. scientific articles or web pages relating to their knowledge or expertise, court cases they were involved in, or their parliamentary interventions, respectively). The aim would be to recommend experts to people who need their services according to their needs.

Formally, let $I=\{i_1,\ldots,i_r\}$ be the set of items to be recommended. Linked with each of these, $i\in I$, is a set of $n_i$ text documents $D^i=\{d^i_1,d^i_2,\ldots,d^i_{n_i}\}$.

Each item will also be represented by a profile that contains in one way or another the content of its related documents (the terms appearing in them). These profiles can basically be organised in one of two ways:

\begin{itemize}
    \item Monolithic profiles: where all the documents linked to each item $i$ are concatenated to create a single document, $d^i=\cup_{j=1}^{n_i} d^i_j$. This macro document will act as a unique profile $p_{Mono}^i$ for item $i$, i.e. $p_{Mono}^i=\{d^i\}$. 
    \item Atomic subprofiles: where for item $i$, its profile will comprise as many subprofiles as documents attached to it but in an isolated, unconcatenated way, i.e. $p_{Atom}^i=\{d^i_1,d^i_2,\ldots,d^i_{n_i}\}$ (each document is treated as a subprofile in itself).
\end{itemize}

\begin{figure}[tb]
\centering
\includegraphics[width=\textwidth]{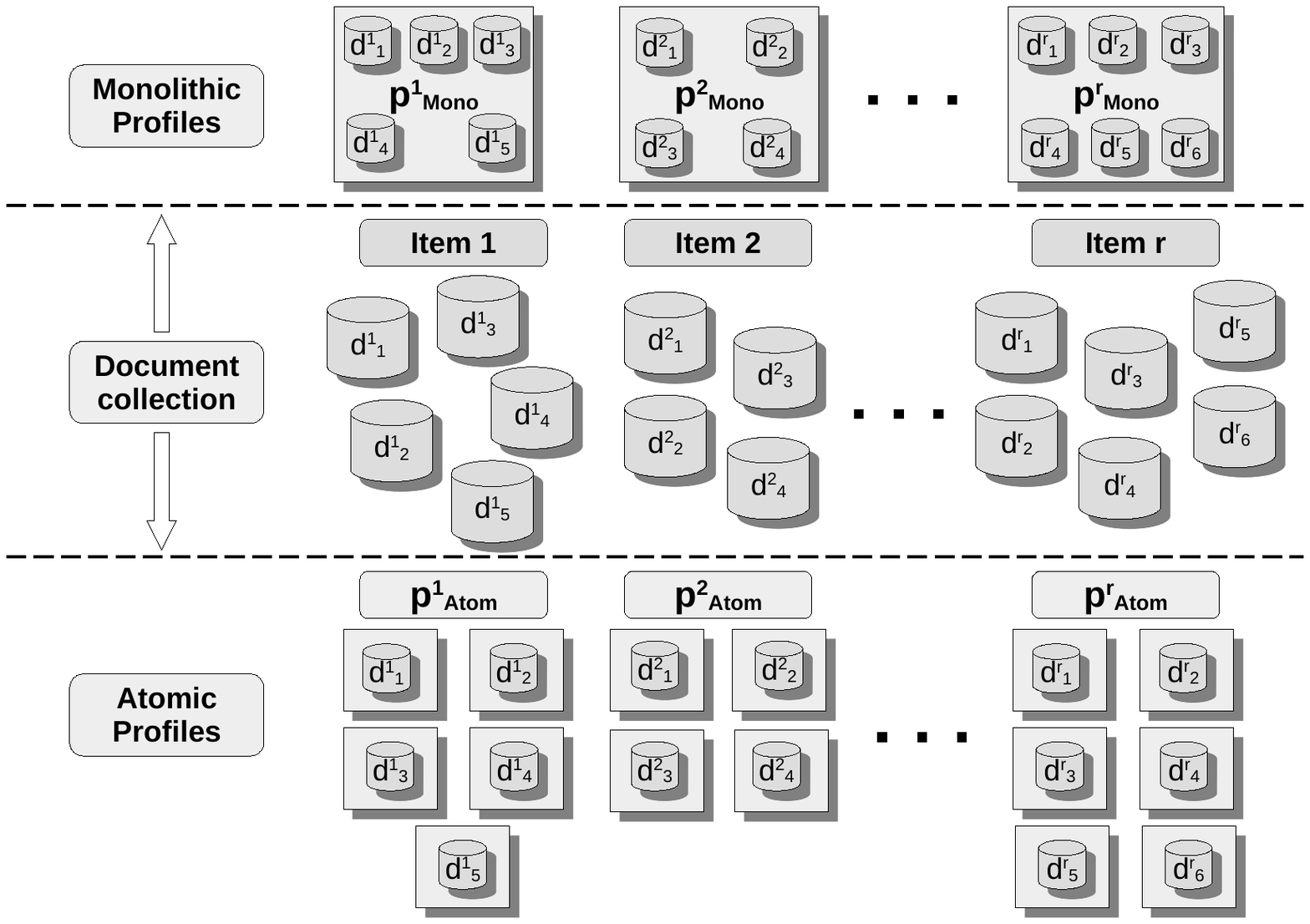}
\caption{Building monolithic and atomic (sub)profiles}
\label{fig:MonoAtomProfiles}
\end{figure}

The collection of items is then represented by a set of profiles, ${\mathcal{P}} = \{p^{i_1},\ldots,p^{i_r}\}$, which will serve as retrieval units in this context of CBR systems. A graphical representation of this profile construction process is shown in Figure \ref{fig:MonoAtomProfiles}.

These two basic ways of classifying profiles as either monolithic or atomic correspond in expert finding literature with the so-called profile-based methods and document-based methods, respectively \cite{Balog12}. There is no general agreement about which method is preferable, and although document-based methods tend to be considered better than profile-based methods, profile-based methods perform better in certain cases \cite{egovis15,LiuCroft05}). We will therefore use both approaches as baselines in our experiments.

We believe that there is room for improvement between the complete atomisation of isolated and homogeneous subprofiles and the extreme compaction of monolithic and heterogeneous profiles, and that we can create less extreme ways of organising the information relating to each item.

\subsection{Topical profiles} \label{subsec:topical}

One first alternative for organising an item's subprofiles (out of the two basic organisational schemes presented in the previous section) would be to build more homogeneous subprofiles around the different concepts or topics which can be identified in the entire collection of text documents associated to the items. The construction of subprofiles from a topical perspective can be based on a partition of the document collection by means of a clustering algorithm which uses the documents' terms as features. This would identify the different clusters of documents according to their subjects, placing all the conceptually-related documents in the same cluster. For an item $i$, each subprofile will correspond to the concatenation of the documents associated to $i$ which are assigned to the same cluster, and this results in a set of topically homogeneous subprofiles. It is apparent that while this clustering process is global in that it is carried out with the entire document collection (not with the documents associated to an item), the subprofile construction is local, as the subprofiles only contain the text of documents associated to the item $i$. 

Although there are many ways of performing this clustering process, in this paper we will use LDA (Latent Dirichlet Allocation) \cite{Blei03} for this purpose. LDA finds $k$ latent topics, $x_1,x_2,\ldots,x_k$, in a document collection\footnote{The number of topics, $k$, is an input parameter of LDA.}, where each topic $x_l$ is characterised by a conditional probability distribution of terms, $p(t \mid x_l)$, and determines for each document $d$ a probability distribution of topics, $p(x_l \mid d)$. For example, in the context of recommending Computer Science journals, we could highlight the fact that the articles published in all the journals deal with five different topics (to clarify the example, these will be called clustering, classification, regression, association, and feature selection). One document might deal only with classification (100\%), a second one might mainly be about feature selection (70\%) but may also discuss classification as a secondary topic (30\%), a third one might mostly cover regression (90\%) but also briefly touch on feature selection (10\%), and a fourth one might cover all the topics equally (20\% each), as it could be an introduction to Machine Learning.

\begin{figure}[tb]
\centering
\includegraphics[width=\textwidth]{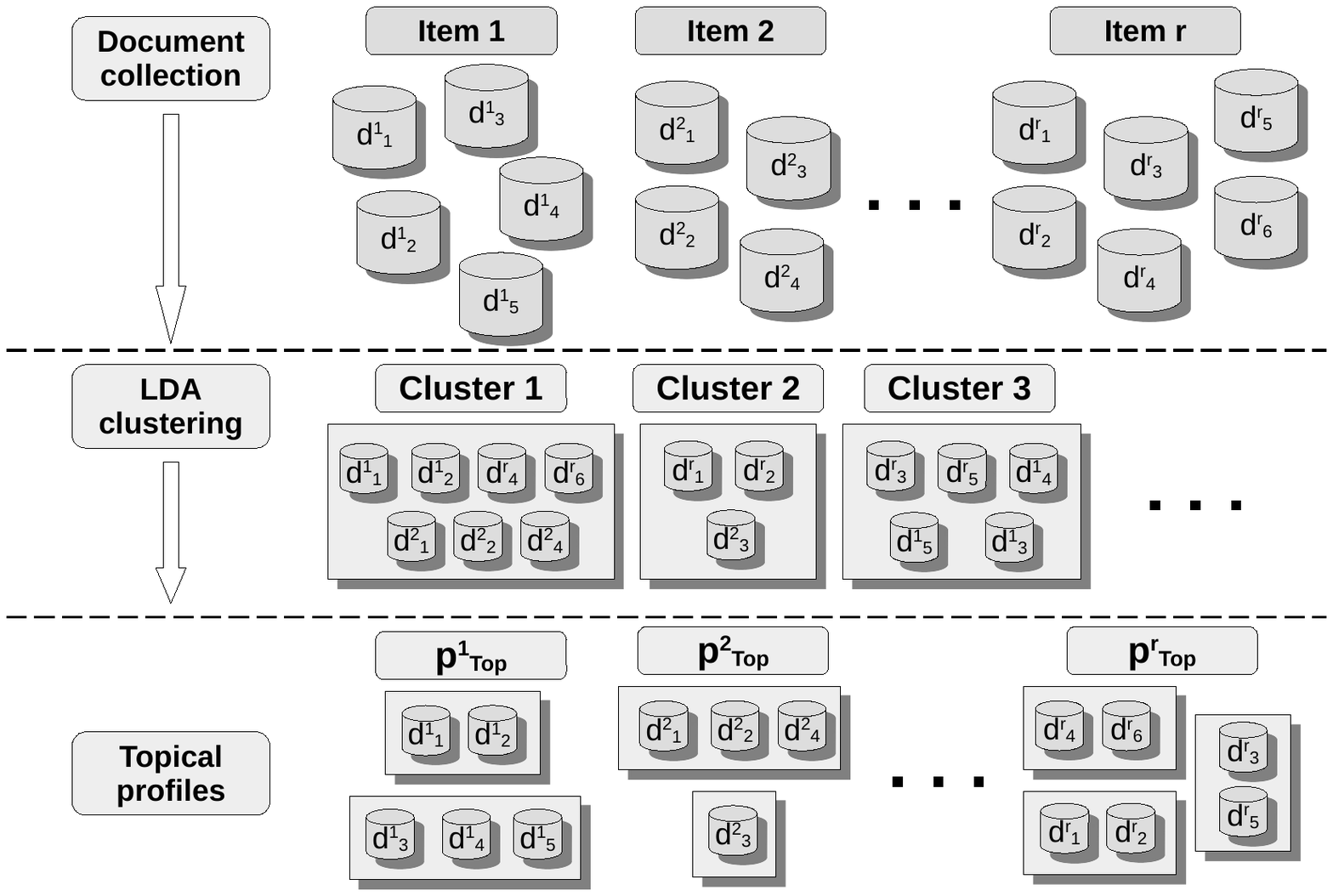}
\caption{Building topical profiles}
\label{fig:TopicalProfiles}
\end{figure}

The clustering generated by LDA obtains $k$ clusters, one for each topic $x_l$, where each document is assigned to its most probable topic. For each item $i$, there are therefore at most\footnote{If an item $i$ has no associated documents about the main topic of $x_l$, then the corresponding subprofile does not exist for that item.} $k$ subprofiles, each containing the concatenated text of those documents associated to $i$ on the main topic of $x_l$. Figure \ref{fig:TopicalProfiles} illustrates the entire process for generating the topical subprofiles.  

In terms of a more formal description, given the set of all the documents ${\mathcal{D}}=\cup_{i=1}^r D^{i}$, each cluster, ${\mathcal{D}}_l$, $l=1,\ldots,k$, consists of the documents of the items which are associated to the $l$-th topic, $x_l$ (documents where the most probable topic is $x_l$), that is to say:
\begin{equation} \label{Dl}
{\mathcal{D}}_l = \{d^{i_j}_m \,\mid\, x_l=\arg\max_{s=1,\ldots,k} p(x_s \mid d^{i_j}_m), \; j=1,\ldots,r, \, m=1,\ldots,n_{i_j}\}
\end{equation}

From these sets of documents relating to each of the $k$ topics, the subprofiles of each item $i$ must be constructed by grouping the documents in each global cluster that are associated to this item $i$, thereby obtaining a local cluster, $D^i_l={\mathcal{D}}_l\cap D^i$. Each item will therefore have assigned as many subprofiles as local clusters have been generated for it. The subprofiles of item $i$ are then documents, $d^{i,l}$, built by concatenating the documents within each local cluster $D^i_l$, $d^{i,l}=\cup_{d^i_j\in D^i_l} d^i_j$. In this case, the topical profile for an item $i$ is $p_{Top}^i=\{d^{i,1}, d^{i,2},\ldots,d^{i,k}\}$.

\subsection{Temporal profiles}

Another alternative for organising item subprofiles assumes that all the documents relating to any item have an assigned date (e.g. a publication date). We could then sort them accordingly and establish certain temporal divisions to exploit the temporal dimension. From each period, homogeneous subprofiles (in a temporal sense) could be built to represent an item by grouping together all the documents associated to that item which belong to the same period. 

For example, in the context of scientific venue recommendation, imagine the situation of a journal about machine learning with an extensive record in the publication of articles. As expected, there are times when there are a large number of papers on the same topic. After a while, new research topics appear and as these are published, they displace and possibly replace some of the existing topics. Initially, the majority of work published in journals of this kind dealt with neural networks but after a few years, researchers became more interested in support vector machines, and nowadays the focus has shifted towards deep learning and this is currently one of the most published topics. Splitting the time line into different intervals and building subprofiles for the corresponding journal in each period is an alternative and simple way of endorsing a temporal perspective to subprofiles and reflects how the focus of the papers has changed over time.

In order to formalise this idea, let us consider the $h+1$ time points $t_0<t_1<\ldots<t_h$ and the $h$ temporal intervals $[t_{u-1},t_u),\, u=1,\ldots,h$. If $date(d)$ is a function that returns the date of document $d$, then the $h$ global temporal clusters ${\mathcal{T}}_u$ are defined as follows:

\begin{equation} \label{Tu}
{\mathcal{T}}_u = \{d^{i_j}_m \,\mid\, t_{u-1}\leq date(d^{i_j}_m)<t_u, \; j=1,\ldots,r, \, m=1,\ldots,n_{i_j}\}.
\end{equation}

The local temporal clusters for each item $i$ are built as in the case of topical clusters by grouping the documents associated to $i$ that belong to each global cluster ${\mathcal{T}}_u$, i.e. $T^i_u={\mathcal{T}}_u\cap D^i$. Each subprofile for item $i$ concatenates the documents in $T^i_u$ to form a single document $d^{i;u}=\cup_{d^i_j\in T^i_u} d^i_j$. Each item, therefore, is now represented by at most $h$ temporal subprofiles\footnote{This is because it is possible for certain items not to have any associated documents in a given period.}. The temporal profile for item $i$ in this case is $p_{Temp}^i=\{d^{i;1}, d^{i;2},\ldots,d^{i;h}\}$.

\subsection{Hybrid profiles: Combining topical and temporal profiles}

The next natural step is to combine both previous ways of constructing profiles by simultaneously exploiting the topical and temporal dimensions in order to obtain  homogeneous subprofiles in terms of these two properties. This homogeneity could be reached in two different ways:

\begin{itemize}
 \item By first discovering the underlying global topics within the whole collection and creating their corresponding clusters (topical division), as explained in Section \ref{subsec:topical}, and secondly by splitting them into temporal units (temporal division), from which the final subprofiles will be constructed. This approach is then a topical-temporal one.
 
 \item By first splitting the collection into temporal units (temporal division), secondly by discovering the underlying topics within each single temporal partition (locally), and finally by building the final subprofiles from them. In this case, this is a temporal-topical approach. It should be noted that in each temporal division, the discovered topics would be different. 
\end{itemize}

More specifically, for the topical-temporal approach, let ${\mathcal{D}}_l$ be defined as in Equation (\ref{Dl}), $l=1,\ldots,k$, and ${\mathcal{T}}_u$ be defined as in Equation (\ref{Tu}), $u=1,\ldots,h$. The global topical-temporal clusters ${\mathcal{DT}}_{lu}$ are then defined as 

\begin{equation} \label{DTlu}
{\mathcal{DT}}_{lu} = {\mathcal{D}}_l \cap {\mathcal{T}}_u, \; l=1,\ldots,k, \; u=1,\ldots,h.
\end{equation}

Given an item $i$, its local clusters are obtained by joining the documents found in each global topical-temporal cluster which are associated to $i$: $DT^i_{lu}={\mathcal{DT}}_{lu} \cap D^i$. As in the previous cases, the documents within each local cluster are concatenated to form the corresponding subprofiles, $d^{i,l;u}=\cup_{d^i_j\in DT^i_{lu}} d^i_j$. The topical-temporal profile for item $i$ is $p^i_{TopTemp}=\{d^{i,1;1},\ldots,d^{i,k;h}\}$.

In the case of the temporal-topical approach, starting from each ${\mathcal{T}}_u$ as defined in Equation (\ref{Tu}), $u=1,\ldots,h$, as the document collection, we use LDA to obtain the $k$ topics $x_{u1},\ldots,x_{uk}$ corresponding to this collection\footnote{It should be noted that we must apply LDA to every $h$ temporal subcollections of documents, thereby obtaining specific topics for each time period.}. We then proceed in the same way as with the topical profiles, i.e. we obtain the clusters ${\mathcal{TD}}_{ul}$ as the subset of documents of ${\mathcal{T}}_u$ where the most probable topic is $x_{ul}$:

\begin{eqnarray} \label{TDul}
{\mathcal{TD}}_{ul} = \{d^{i_j}_m\in {\mathcal{T}}_u \,\mid\, x_{ul}=\arg\max_{s=1,\ldots,k} p(x_{us} \mid d^{i_j}_m), \; j=1,\ldots,r, \, m=1,\ldots,n_{i_j}\}, \\ \nonumber
\; l=1,\ldots,k, \; u=1,\ldots,h.
\end{eqnarray}

We then obtain the clusters associated to each item $i$ as $TD^i_{ul}={\mathcal{TD}}_{ul} \cap D^i$. Finally, the documents in $TD^i_{ul}$ are concatenated to build the subprofiles, $d^{i;u,l}=\cup_{d^i_j\in TD^i_{ul}} d^i_j$. The temporal-topical profile for item $i$ is $p^i_{TempTop}=\{d^{i;1,1},\ldots,d^{i;h,k}\}$.

It is worth noting that in both cases in this process, each item $i$ will have associated at most $k*h$ subprofiles because there might not be any document associated to $i$ which deals with a specific topic at a given time period.

\section{Evaluation and results} \label{evaluation}
In this section, we will detail everything relating to the evaluation of the previously explained alternatives for item profiling and also the results obtained. 

The context on which this experimentation is based is publication venue recommendation. Given an article (or at least an abstract, title and keywords), the problem is to recommend to the active user the most suitable venue for publishing such a paper on account of the suitability of the scope of the journal. In this context, the items to be recommended are journals and the text documents associated to the items are the articles published in these journals.

The following sections will present all the details of the experimental design and also the results of the experiments. 

\subsection{Test collection}

The test collection used in the experimentation is called PMSC-UGR \cite{Albusac18} and has been created by the authors from PubMed and Scopus. It originally  contained $762,508$ articles from $12,396$ journals in the biomedical domain, with a title, abstract, keywords, citations and authors for each paper. Out of all the authors from this selection of papers, those who were unequivocally represented by their corresponding ORCID codes were finally selected, leaving a total of $20,406$ authors.

\subsection{Recommendation model}

Once all the (sub)profiles\footnote{All of the methods presented (with the exception of the one based on monolithic profiles) generate subprofiles, i.e. several subprofiles per item. The monolithic approach builds exactly one profile per item and this is why we write (sub)profiles.} relating to each item (journal) have been built, the IR technique is used to make content-based recommendations, regardless of the method applied. First, all these (sub)profiles\footnote{It should be remembered that each subprofile is a text document.} are indexed by an IR system. In our case, we have developed an indexing programme based on the Lucene library\footnote{https://lucene.apache.org/}. Previously, this piece of software removes stop words and performs stemming, indexing only the resulting stems.

Given an active user represented by any kind of textual source, the objective is to obtain a ranking of relevant items. Considering that CBR is based on IR, an IR model will be responsible for retrieving relevant items in respect of the user information need and this serves as a query for the IRS. More specifically, the textual content of a paper to be published by the user represents a query submitted to the IRS. By using the Lucene implementation of the Language Model (with Jelinek-Mercer smoothing), a retrieval programme computes a ranking of journal (sub)profiles which are sorted decreasingly according to their relevance to the query.

As mentioned in the previous paragraph, the ranking consists of subprofiles. However, since the active user requires journal recommendations, the subprofile ranking should then be transformed into a journal ranking. For this purpose, a final fusion process combines the scores of the subprofiles for each journal and generates a final journal ranking. This is performed by means of the {\it{CombLgDCS}} fusion method \cite{deCampos17b}, which aggregates the scores of all of the journal subprofiles, decreasing them proportionally to the logarithm of their positions in the ranking. This fusion is not necessary for monolithic profiles since in this case there is only one profile per journal. 

\subsection{Discovering latent topics}

% In order to discover the latent topics in the test collection and obtain topically homogeneous subprofiles, we have used a Python implementation of LDA given in the Gensim library, with the default values for hyper-parameters $\alpha$, which controls the prior distribution over the topic weights in each document, and establishes a fixed symmetric prior of 1/$k$ (where $k$ is the number of topics), and $\beta$ is the parameter for the prior distribution over word weights in each topic and which is also set to the default value of 1/$k$.

During this process, terms that appear in fewer than $750$ documents are ignored and also those that appear in more than $90\%$. Finally, only a maximum of 5000 of the most frequent terms in the corpus have been considered as LDA input. These figures correspond to previous experiments to set the most suitable size of the vocabulary. Nevertheless, the subprofiles built with this technique contain all the terms from the original documents.

\subsection{Baselines}

In this research, two methods of building profiles are considered as baselines: monolithic profiles and atomic subprofiles. Not only are these the simplest ways of organising the textual information without taking into account topical and temporal features, they are also the most extreme alternatives whereby all the articles in the journal comprise the same profile (with only one profile per journal) or each individual article forms a subprofile (there will be as many subprofiles as articles). It is therefore hoped that these more complex ways of structuring profiles can improve the quality of CBR.

\subsection{Experimental design}

For evaluation purposes, we have restricted the PMSC-UGR collection to those papers published between 2007 and 2016 which appear in journals with more than 100 papers in this period\footnote{We have removed the journal PlosOne from this set because it has a much greater number of papers than the others.}, leaving the dataset with a total of $1002$ journals. The collection has then been split into two partitions: the first comprises articles dating from 2007 to 2015 (a total of $276,679$ papers) and this will be reserved for building the (sub)profiles and serve as the training set; and the second only contains articles from 2016 ($32,864$ articles) and this will be used as the test set. This holdout method is suitable for this evaluation and does not require cross-validation to obtain reliable results given the large number of articles in the test set. Each article from the test partition will be considered as a query to be submitted to the underlying IRS and this query will comprise the combined text of its title, abstract and keywords.

Starting from the training partition, the following types of (sub)profiles will be considered for our experiments:

\begin{itemize}
 \item Monolithic profiles ($Mono$)
 \item Atomic subprofiles ($Atomic$)
 \item Topical subprofiles ($Top$)
 \item Temporal subprofiles ($Temp$): These are built from four temporal partitions of two years each (2007--2008, 2009--2010, 2011--2012 and 2013--2014) and one of only one year (2015), i.e. $h=5$.
 \item Topical + Temporal ($TopTemp$): After applying a global LDA to the entire article collection, subprofiles are built in the five temporal partitions.
 \item Temporal + Topical ($TempTop$): Subprofiles are built from the five temporal partitions after applying local LDA to the article collection in each partition.

\end{itemize}

In addition, $10$ instances of randomised subprofiles have been created ($Random1$, $\dots$, $Random10$). Each comprises $5$ random partitions of the articles from the training set, thereby replacing the 5 temporal subprofiles with 5 random subprofiles. The retrieval effectiveness of these data sets will also be measured in order to test whether the temporal divisions differ from randomness. 

For the LDA algorithm on which the $Top$, $TopTemp$ and $TempTop$ approaches are based, it is necessary to set up the $k$ parameter, i.e. the number of latent topics to be discovered. This is not an easy task because the quality of the results could be very high depending on this value. In this research, we have considered three different values, two of which are related to medical categories or specialities:

\begin{itemize}
 \item number of comprehensive medical specialities extracted from the Medical School blog at St George's University\footnote{https://www.sgu.edu/blog/medical/ultimate-list-of-medical-specialities}, $k=20$
 \item number of second-level categories of the MESH thesaurus\footnote{https://meshb.nlm.nih.gov/treeView}, $k=110$
 \item  $k=400$, in order to test a very large number of topics
\end{itemize}

It is apparent that these three values attempt to cover a wide range of topics, from a low to a relatively large number of them, in order to evaluate the performance of the different types of profiles according to the number of topics discovered by the LDA algorithm. The underlying idea is also to choose meaningful values of $k$ relating to medical categories. 

For the truly temporal approaches, i.e. $Temp$, $TopTemp$ and $TempTop$, and for $Atomic$, which also supports a temporal treatment, a temporal decay might also be considered to reduce the influence of older articles or subprofiles. This is a complementary way of introducing the temporal dimension into this journal recommendation problem. The underlying idea is that a user would be more interested in journals where their last published articles are closer to the paper to be published rather than ``older'' articles. In order to implement this idea, once a query has been submitted (the text of the article) and the score or Relevance Status Value (RSV) for each article or subprofile has been computed, a decay function is applied in order to modify the corresponding RSV according to the temporal distance to 2015, which is the year of the ``newest'' papers in our collection. Several decay functions might be found in the literature, for example those presented in \cite{LTPGN15}, but in our case, two have been implemented to be tested:

\begin{itemize}
    \item linear: $RSV_{LinearD} = \frac{NRSV}{1 + Penalty}$ 
    \item doubled squared root: $RSV_{2SqrtD} = \frac{NRSV}{\sqrt{\sqrt{1 + Penalty}}}$
\end{itemize}

\noindent where $NRSV$ is the normalised $RSV$, computed by dividing the corresponding $RSV$ by the maximum of the ranking and $Penalty$ = $2015$ - $Publication$ $year$. For individual articles in the atomic subprofiles, their publication year is used directly to compute the new score, whereas for the remaining temporal subprofiles, their average value is incorporated into the decay formulas since they contain articles over two years.

Although other decay functions were implemented and tested in this study, these two were selected in order to show two different penalisation behaviours as it is evident that the doubled squared root decay penalises older articles more smoothly than the linear one. The other functions presented behaved in a similar or even worse way and that is why these two are included as representative.

A third value of the decay parameter would be "None", which means that no temporal penalisation is considered in the RSV.

\subsection{Evaluation measures}

In order to evaluate the recommendation, and before presenting the evaluation measures, it is important to determine the ground truth: in this case, only one journal is relevant for each query (test article) and this is the journal where the paper has actually been published. 

The following evaluation measures found in the literature to determine the quality of the results obtained by venue recommendation methods are the most common: 

\begin{itemize}
 \item Recall@X (R@X): This measures the ability of recommending the relevant journal where the test article has been published in the first $X$ journals in the ranking (recommended venues). In other words, we compute the average number of times where the actual venue where a test paper was published is among the first $X$ recommended venues. In previous work on venue recommendation (e.g. \cite{Luong12, Medvet14, Wang18}), this measure is also called accuracy@X. Two thresholds $X$ are considered to compute the values of this measure: $1$ and $5$. $X=1$ is considered because the number of relevant journals is $1$ and so we would like to know how successful it would be to recommend only one journal. Since the user does not usually obtain only one recommendation but a number of alternative journals, $X=5$ is also used as the threshold.
 \item Mean Reciprocal Rank (MRR@Y): In this case, the idea is to reflect how high in the ranking the only relevant journal is recommended. It therefore computes the average of the inverse of the positions in the ranking for the journal where each test paper was published. The total number of results (journals) in the ranking is limited to $Y$ \footnote{If the relevant journal is not within the first $Y$ journals in the ranking, then the value is 0.}. We have only considered the top 40 positions in the ranking, i.e. $Y=40$.
 In our specific case where only an item (a journal) is relevant, this measure coincides with mean Average Precision, MAP@Y.
\end{itemize}

\subsection{Results}

\begin{table}[htb]
\begin{center}
\caption{Results of the experiments (the best results for each column are shown in bold)}
\begin{tabular}{cccccc}
\toprule
\textbf{Subprofiles} & \textbf{\#Topics} & \textbf{Decay} & \textbf{R@1} & \textbf{R@5} & \textbf{MRR@40}  \\ 
\midrule
TempTop & 110 & 2Sqrt & \textbf{0.2501} & \textbf{0.5655} & \textbf{0.3935} \\
TempTop & 20 & 2Sqrt & 0.2490 & 0.5619 & 0.3923 \\ 
TempTop & 20 & None & 0.2466 & 0.5615 & 0.3903 \\ 
TempTop & 400 & 2Sqrt & 0.2455 & 0.5596 & 0.3883 \\ 
TempTop & 110 & None & 0.2451 & 0.5638 & 0.3898 \\ 
TempTop & 400 & None & 0.2434 & 0.5575 & 0.3860 \\ 
TopTemp & 400 & 2Sqrt & 0.2373 & 0.5462 & 0.3783 \\ 
TopTemp & 110 & 2Sqrt & 0.2368 & 0.5463 & 0.3781 \\ 
Top& 20 & None & 0.2348 & 0.5400 & 0.3760 \\ 
Top & 110 & None & 0.2346 & 0.5448 & 0.3763 \\ 
Top & 400 & None & 0.2345 & 0.5462 & 0.3771 \\ 
Temp &  & 2Sqrt & 0.2341 & 0.5431 & 0.3758 \\ 
TempTop & 110 & Linear & 0.2335 & 0.5307 & 0.3692 \\ 
Atomic &  & 2Sqrt & 0.2331 & 0.5403 & 0.3731 \\ 
TopTemp & 110 & None & 0.2331 & 0.5431 & 0.3747 \\ 
TopTemp & 20 & 2Sqrt & 0.2329 & 0.5419 & 0.3746 \\ 
TopTemp & 400 & None & 0.2328 & 0.5430 & 0.3746 \\ 
TempTop & 400 & Linear & 0.2308 & 0.5266 & 0.3651 \\ 
TopTemp & 400 & Linear & 0.2298 & 0.5179 & 0.3615 \\ 
TempTop & 20 & Linear & 0.2294 & 0.5267 & 0.3657 \\ 
TopTemp & 110 & Linear & 0.2290 & 0.5207 & 0.3618 \\ 
TopTemp & 20 & Linear & 0.2286 & 0.5230 & 0.3629 \\ 
TopTemp & 20 & None & 0.2286 & 0.5361 & 0.3699 \\ 
Atomic &  & None & 0.2282 & 0.5370 & 0.3696 \\ 
Temp &  & None & 0.2258 & 0.5330 & 0.3671 \\ 
Atomic &  & Linear & 0.2255 & 0.5118 & 0.3558 \\ 
Random2 &  & None & 0.2245 & 0.5340 & 0.3656 \\ 
Random1 &  & None & 0.2243 & 0.5326 & 0.3658 \\ 
Random6 &  & None & 0.2242 & 0.5330 & 0.3660 \\ 
Random9 &  & None & 0.2241 & 0.5321 & 0.3653 \\ 
Random3 &  & None & 0.2240 & 0.5319 & 0.3657 \\ 
Random5 &  & None & 0.2239 & 0.5330 & 0.3656 \\ 
Random4 &  & None & 0.2239 & 0.5333 & 0.3655 \\ 
Random8 &  & None & 0.2238 & 0.5332 & 0.3655 \\ 
Random0 &  & None & 0.2236 & 0.5335 & 0.3636 \\ 
Random0 &  & None & 0.2236 & 0.5335 & 0.3653 \\ 
Monolithic &  & None & 0.2236 & 0.5278 & 0.3653 \\ 
Random7 &  & None & 0.2227 & 0.5314 & 0.3648 \\ 
Temp &  & Linear & 0.2147 & 0.5159 & 0.3543 \\ 
\botrule
\end{tabular}
\label{tab:results}
\end{center}
\end{table}

The results of our experiments are displayed in Table \ref{tab:results}.
Although the rankings of methods obtained for the different performance metrics are not identical, the trends are the same. In fact, if we calculate the Pearson correlation between these rankings, we always obtain correlation coefficients which are greater than 0.86 (if we exclude the results obtained by the random subprofiles and also those which use linear decay, which are both quite poor, then all the correlation coefficients are greater than 0.98). We can, therefore, comment on our results without referring to any specific metric.

If we first focus on RQ1, we can conclude that the use of the temporal dimension alone ($Temp$ subprofiles, without decay) only very slightly improves the results obtained by the $Mono$ baseline. This implies that the impact of dividing the monolithic profiles into several parts which are only based on temporal criteria is limited. In addition, the $Atomic$ approach performs almost the same as $Temp$. Splitting the collection into years and building the subprofiles upon them does not, therefore, offer any clear advantage. The random subprofiles, Random$i$ (with the same number of subprofiles as $Temp$), perform almost identically to $Mono$ and  worse than $Temp$. This implies that the temporal dimension has a slightly positive effect on the results which is not attributable merely to the fact of creating several subprofiles.

When a decay factor is also used, very poor results are obtained for $Temp$ (even worse than the baselines) with the linear version and much better results with the doubled squared root version (2Sqrt) (which is smoother than linear). Consequently, the performance of $Temp$ clearly depends on the aid that decay functions can provide. This behaviour of the three decay methods (Linear $<$ None $<$ 2Sqrt) also persists when the temporal dimension is combined with the topical dimension\footnote{Although in this case, the difference between None and 2Sqrt is smaller.} or when it is applied to the other baseline, $Atomic$, so that the preferred version of decay is always 2Sqrt. $Temp$ is also better than $Atomic$ (using 2Sqrt decay). By way of conclusion, a well-designed decay function incorporated into the recommendation process can boost performance, and this therefore answers RQ2.
 
In terms of RQ3, the use of topical subprofiles ($Top$) clearly improves the results of the baselines $Mono$ and $Atomic$, regardless of the number of topics selected. Since there are no very important differences between the results of $Top$ with a different number of topics, this parameter does not seem critical for good behaviour. $Top$ is also better than $Temp$ without decay and has a similar performance to $Temp$ with 2Sqrt. However, the contributions of $Top$ and $Temp$ to  improved performance seem to be based on different premises, so that their combination could generate a kind of synergy. This is indeed the case but it depends on the way the topical and temporal subprofiles are combined. When $TopTemp$ subprofiles are used, i.e. first a topical division and then the temporal division (and using the same topics in every time period), we do not observe any clear improvement in the results in terms of using only either $Top$ or $Temp$ (and only a miniscule improvement is obtained). However, when we use the other proposed combination of $TempTop$ subprofiles, i.e. first a temporal division and then the topical division (with the topics being specific for each time period), a clear improvement is apparent. Moreover, the $TempTop$ results are always the best ones for all the metrics, regardless of the number of topics being considered. We think that this is due to the fact that topic identification, and the subsequent construction of subprofiles, is tailored to the set of articles included in each temporal partition, which is more precise and totally adapted to the content of such articles. Additionally, the $TopTemp$ approach is more general and not so well fitted to the texts in each temporal partition. Hybridising topical and temporal profiles is, therefore, a very interesting approach but only if temporal divisions are made and subprofiles built based on topic discovery in each time-based partition (RQ4). The absolutely best results are obtained using $TempTop$ subprofiles with 110 topics and 2Sqrt decay (RQ5).

Finally, and in order to try to verify these conclusions, a statistical significance test has been applied for the measure $R@1$. More specifically, the McNemar test \cite{McNemar47} was selected, which is a non-parametric test for paired data, as recommended in \cite{Dietterich98} for comparing machine learning algorithms. It has been run for families of subprofiles (for the three values of $k$ and the best decay method, when applicable): $TempTop$ with 2Sqrt, $TopTemp$ with 2Sqrt and $Top$ without decay. The idea is to first determine if there are significant differences in each family. The results of these tests fulfil the same pattern for $TempTop$ and $TopTemp$ families: there are no differences between the two top $k$ values and there is with the worst. In the case of $TempTop$, there are no differences between $110$ and $20$, but there are with $400$. For $TopTemp$, there are no differences between $400$ and $110$, but there are for $20$. In terms of $Top$, there are no differences between the three values of $k$. A final series of tests is run between the best values of each family, including in this case the baselines: $TempTop$, 110, 2Sqrt; $TopTemp$, 400, 2Sqrt; $Top$, 20, None; $Temp$, 2Sqrt; $Atomic$, 2Sqrt; and $Mono$, None. The results show that there are significant differences between $TempTop$ and the others and also between $Mono$ and the others, and there are no differences between $TopTemp$, $Top$ and $Temp$. And there are differences between $TopTemp$ and $Atomic$, but not between $Atomic$, $Top$ and $Temp$. By way of summary, $TempTop$ is clearly the best option for combining temporal and topical dimensions and $Mono$ is the worst alternative.

\section{Conclusions and Further Research} \label{conclusions}

In this paper, we have focused on testing how useful temporality, topicality and the combination of these are for the problem of building and using profiles in content-based recommendation. We have proposed two different ways of hybridisation. A biomedical journal collection on publication venue recommendation was used to test our findings and these revealed that the combination of these two types of approaches is a good alternative although it is important to note that order matters in terms of performance. From our experiments, we can conclude that the best option is to create temporal partitions and discover the latent topics starting from the papers in each partition using LDA, and then to construct the profiles. It is important to mention that the number of topics for building the profiles is not a critical parameter. The application of a decay factor might be a valuable aid but it clearly depends on the quality of the penalising function. In our context, 2Sqrt helps to improve the performance of the recommendation with hybrid subprofiles. 

In terms of future lines of research, we plan to explore other methods of combining temporal and topical dimensions to obtain better subprofiles. One alternative is the use of temporal topic models \cite{Blei06,DRB19}. Another option is to use methods based on the aggregation or fusion \cite{Wu12} of topical and temporal rankings individually obtained by $Top$ and $Temp$, respectively. Another research line is to design high quality decay functions that boost the performance of hybridisation. Finally, we also plan to study the most suitable ways of explaining the recommendations \cite{Tintarev12} offered by our models.

\subsubsection*{Acknowledgements}
This work has been co-funded by the Spanish Ministerio de Econom\'ia y Competitividad under project PID2019-106758GB-C31, the Junta de Andalucía and University of Granada under project A‐TIC‐146‐UGR20 (Programa Operativo FEDER Andalucía 2014-2200), and the European Regional Development Fund (ERDF-FEDER).

\section{Statements and Declarations}
\subsubsection*{Funding}

This study was funded by the Spanish Ministerio de Econom\'ia y Competitividad under project PID2019-106758GB-C31, the Junta de Andalucía and University of Granada under project A‐TIC‐146‐UGR20 (Programa Operativo FEDER Andalucía 2014-2200), and the European Regional Development Fund (ERDF-FEDER).

\subsubsection*{Competing interests}

The authors have no relevant financial or non-financial interests to disclose.

\subsubsection*{Author contribution}

All authors contributed to the study conception and design. Material preparation, programming, data collection and analysis were performed by the three of us. The first draft of the manuscript was written collaboratively and all authors commented on previous versions of the manuscript. All authors read and approved the final manuscript.

\end{document}